\documentclass{npw}
\usepackage{graphicx}

\usepackage{color}

\usepackage{amsmath}
\usepackage{amssymb}
\usepackage{amsfonts}
\usepackage{graphicx}
\usepackage{tabularx}
\usepackage{multirow}
\usepackage{lscape}
\usepackage{rotating}
\usepackage{bbm}

\allowdisplaybreaks

\renewcommand{\vec}[1]{{\mathbf{#1}}}
\newcommand{\vsigma}{\boldsymbol{\mathbf\sigma}}
\newcommand{\vnabla}{\boldsymbol{\mathbf\nabla}}
\newcommand{\p}[3]{\hat{P}^{#1}_{#2#3}}

\begin{document}

\title{Skyrme pseudo-potential-based EDF parametrisation 
       for spuriousity-free MR-EDF calculations}

\author{
 J. Sadoudi\email{sadoudi@cenbg.in2p3.fr} \,  and M. Bender \\
  {\it Univ. Bordeaux, CENBG, UMR 5797, F-33170 Gradignan, France.} \\
  {\it CNRS/IN2P3, CENBG, UMR 5797, F-33170 Gradignan, France. } \\
  K. Bennaceur, 
  D. Davesne and 
  R. Jodon\\
  {\it Universit\'e de Lyon, F-69003 Lyon, France; } \\
  {\it Universit\'e Lyon 1,
      43 Bd. du 11 Novembre 1918, F-69622 Villeurbanne cedex, France}\\
  {\it CNRS-IN2P3, UMR 5822, Institut de Physique Nucl{\'e}aire de Lyon } \\
T. Duguet \\ 
  {\it CEA-Saclay DSM/Irfu/SPhN, F-91191 Gif sur Yvette Cedex, France, and }\\
  {\it National Superconducting Cyclotron Laboratory and Department of Physics and Astronomy,} \\
  {\it Michigan State University, East Lansing, MI 48824, USA}
}

\date{13 December 2012}

\maketitle

\begin{abstract}
First exploratory steps towards a pseudo-potential-based Skyrme energy 
density functional for spuriousity-free multi-reference 
calculations are presented. A qualitatively acceptable fit can be 
accomplished by adding simple three- and four-body contact terms to the 
standard central plus spin-orbit two-body terms. To achieve quantitative
predictive power, higher-order terms, e.g.\ velocity-dependent three-body 
terms, will be required. 
\end{abstract}

%
\section{Introduction}

Methods based on the use of  effective energy density functionals (EDFs) 
\cite{bender03} currently provide the only set of theoretical tools that 
can be applied to all bound atomic nuclei but the lightest ones in a 
systematic manner irrespective of their mass, isospin, and deformation.
One underlying assumption is that of a universal effective energy 
functional that depends on normal and anomalous one-body density
matrices and that re-sums the in-medium correlations whose details are 
irrelevant for low-energy nuclear structure physics. 
Nuclear EDF methods coexist on two distinct levels. On the first level, 
that is traditionally called self-consistent mean-field theory, 
Hartree-Fock (HF) or Hartree-Fock-Bogoliubov (HFB), a single product 
state provides the normal and anomalous density matrices that enter 
the EDF. We call this type of method a single-reference (SR) 
approach. Exploiting the concept of symmetry breaking as a consequence 
of the nuclear Jahn-Teller effect \cite{Rei84a}, collective correlations 
that correspond to multipole deformations or superfluidity can be easily 
modelled at the expense of losing good quantum numbers. On the second 
level, often called "beyond-mean-field methods", symmetry restoration 
and configuration mixing can be achieved within the Generator Coordinate 
Method (GCM) framework. Such method is referred to as a 
multi-reference (MR) approach \cite{Ben08LH}. The MR techniques aim at 
the explicit description of correlations that 
are related to the small finite size of the nucleus and that neither 
can be easily absorbed into the EDF, nor be described by a single SR 
state. Typical examples are the mixing of different shapes that coexist 
in a nucleus or the restoration of symmetries that are broken at the SR level. 
Besides describing correlated ground states, the MR approach also gives 
access to excited states and transition moments between them taking
selection rules into account. 

Multi-reference techniques based on energy density functionals 
rely on the extension of the EDF to non-diagonal energy kernels. 
The generalised Wick theorem \cite{Ring80a} provides the
formal framework when the EDF is defined as the expectation value of 
a genuine Hamilton operator. The energy functionals widely used in 
nuclear physics, however, are not of this form. We denote these as 
"general functional" in what follows. The standard procedure
for the MR extension of such general functionals is made by formal 
analogy with the Hamiltonian case, cf.\ \cite{Dob07a,I} and references
therein. It has been pointed out a few years ago that the usual 
procedure to set up the non-diagonal kernels of a general EDF is 
ill-defined \cite{Dob07a,I,Ang01a,II,III}. It gives rise to 
spurious contributions to the energy that can manifest themselves 
for example through divergences and/or finite steps when plotting 
the symmetry-restored energy as a function of a collective coordinate
\cite{Dob07a,II}. It also makes MR calculations return non-zero energies 
when restoring \textit{negative} particle numbers~\cite{II,duguetsym}. 
These difficulties can be traced back to a 
breaking of the Pauli principle when setting up the EDF. In one way 
or the other, this is the case for all modern parametrisations of 
the nuclear EDF. It is motivated either by phenomenology or for 
computational reasons \cite{bender03,Was12a}. 

One popular choice for a general nuclear EDF is historically based
on two- and three-body contact pseudo-potentials proposed by Skyrme 
\cite{Skyrme}. In the earliest adjustments of its coupling constants for
nuclear EDF calculations it was used as a genuine Hamilton 
operator\footnote{In the present context, so-called "density-dependent 
effective interactions" do not qualify as genuine Hamilton operators 
or "pseudo-potentials".}. 
This gave rise to three major problems in the resulting parametrisations: 
(i) the incompressibility $K_\infty$ of homogeneous symmetric 
infinite nuclear matter (INM) was unavoidably much larger than the 
empirical one; (ii) fits that constrain other 
INM properties such as isoscalar effective mass $m^*_0/m$, 
saturation density $\rho_{\text{sat}}$ and symmetry energy coefficient 
$a_{\text{sym}}$ to empirical values provided parametrisations
that were unstable in the spin channel as indicated by the
Landau parameter $g_0$ being smaller than $-1$ \cite{Bac75a}, and
(iii) pairing matrix elements were far too small or even repulsive.
Problem (ii) could be circumvented reinterpreting the contact three-body
force $t_3 \, \delta(\vec{r}_1-\vec{r}_2) \delta(\vec{r}_1-\vec{r}_3)$
as a density-dependent two-body force 
$\tfrac{1}{6}\, t_3 \, ( 1 + x_3 \hat{P}^\sigma_{12}) \,
\rho_0^\alpha[(\vec{r}_1+\vec{r}_2)/2] \,
\delta (\vec{r}_1-\vec{r}_2)$ with \mbox{$\alpha = 1$},
where $\rho_0(\vec{r}) = \rho_n(\vec{r}) + \rho_p(\vec{r})$ is the 
isoscalar density and $\hat{P}^\sigma_{12}$ the spin-exchange operator, 
respectively, which alters the so-called
"time-odd" part of the functional that governs the effective spin-spin
interaction. Problem (i) could then be solved by reducing the exponent
$\alpha$ from 1 to values between about $1/6$ and $1/3$. And for most
parametrisations of the Skyrme EDF, problem (iii) was removed 
by using different vertices in the particle-hole and 
particle-particle channels that were independently adjusted to data.
In addition, some specific terms in the EDF are often 
suppressed or approximated to simplify the numerical 
treatment~\cite{bender03}. In the past this has led to high-precision 
parametrisations of the Skyrme EDF that kept a relatively simple 
form~\cite{bender03,chabanat97} and were numerically efficient, at the 
expense of sacrificing the exchange symmetry of the EDF. 
Other variants of the nuclear EDF have been set-up directly through
their energy density without any reference to an underlying effective 
interaction~\cite{Fay00a,Bal08a}.

The breaking of the exchange symmetry that results from this practice 
introduces what is known as spurious "self-interaction" 
in density functional theory for condensed matter~\cite{Per81a}. 
An analysis of the same problem in the context of effective interactions 
as used in nuclear physics has been given in Ref.~\cite{Str78a}. 
A similar spurious "self-pairing" appears when the normal and pairing
parts of the EDF are not derived from the same effective
interaction as pointed out in Ref.~\cite{II}. According to Ref.~\cite{I},
divergences and steps in MR EDF calculations originate from unphysical 
weights such self-interaction and self-pairing contributions are multiplied 
with. Based on this analysis, a regularisation scheme that modifies those 
ill-defined weights while keeping the self-interaction itself
has been proposed in Ref.~\cite{I} and applied with success to the 
case of pure particle-number projection in Ref.~\cite{II}. 

Not all general functionals are regularizable, though. The formalism 
can be applied only to functionals that correspond to polynomials in 
the density matrices \cite{I,III}. Parametrisations of a general 
regularizable Skyrme-type EDF of minimal form have been constructed 
recently \cite{Was12a} and are currently used to test the regularised MR
EDF scheme for general configuration mixing. Constraining the 
EDF to polynomial form makes its fit more difficult.

In parallel, we also started the construction of functionals that are 
free of spuriosities from the outset by setting them up as the expectation 
value of a genuine Skyrme-type Hamilton operator taking all exchange 
and pairing terms into account such that the Pauli principle is obeyed. 
By construction, this removes all spurious contributions to the EDF at 
the price of having less independent terms in the EDF. In what follows, 
we denote such functionals a "pseudo-potential generated EDF".
This replaces the problem of how to set up a formalism for the extension
of a general and flexible EDF to non-diagonal kernels for MR calculations 
by the problems of what is the most efficient form for a predictive 
pseudo potential that can be straightforwardly used in MR calculations 
and how to adjust it to data. This is a difficult task, as a 
pseudo-potential-based functional has much less independent
coupling constants than a general functional of the same form. To be usable 
in SR and MR calculations incorporating pairing correlations, it is 
mandatory that the functional gives a reasonable description of the 
spin and pairing channels of the interaction. Otherwise the level sequence 
after angular-momentum projection is likely to be unrealistic. However,
the difficulty to have even the correct sign of the interaction 
in these two channels has been among of the major motivations to abandon 
Skyrme-type pseudo-potential-based EDFs in favour of the general ones 
in the 1970s.

The aim of this contribution is to present the first explorative steps 
towards the construction of a predictive pseudo-potential based EDF. 
We present the adjustment of a parametrisation that achieves an acceptable
qualitative description in all channels of the interaction by adding 
three- and four-body contact terms without gradients to a standard two-body
Skyrme operator. It will serve to benchmark MR EDF calculations and as 
a reference for what can be achieved without introducing derivatives 
in the new terms.

\section{Pseudo-potential-based EDF}

The pseudo-Hamiltonian used in the present work takes the form
\begin{equation}
\label{eq:Skyrme_int:hamiltonian}
\hat{H}
= \hat{T}^{(1)}
   + \hat{v}^{(2\text{Sk})} + \hat{v}^{(2\text{C})} 
   + \hat{v}^{(3\text{Sk})}
   + \hat{v}^{(4\text{Sk})} \, ,
\end{equation}
where $\hat{T}^{(1)}$ is the kinetic energy operator and
$\hat{v}^{(2\text{C})}$ the Coulomb interaction, which take their standard
form. The $\hat{v}^{(N\text{Sk})}$ are the $N$-body parts of the 
Skyrme-type pseudo-potential for which we consider here the form
\begin{subequations}
\label{eq:Skyrme_int:pseudo_potential}
\begin{align}
\hat{v}^{(2Sk)} 
&=
{t_0} \left( 1 + {x_0} \p \sigma12 \right) \hat{\delta}_{r^{\,}_1r^{\,}_2}   
  \\
&
+ \frac{{t_1}}{2} \left( 1 + {x_1} \p \sigma12 \right) 
  \left( \hat{\vec{k}}^{\,\prime \, 2}_{12} \hat{\delta}_{r^{\,}_1r^{\,}_2} 
        + \hat{\delta}_{r^{\,}_1r^{\,}_2} \hat{\vec{k}}^{\,2}_{12} \right)   
  \\
&
+ t_2 \left( 1 + {x_2} \p \sigma12 \right) \hat{\vec{k}}^{\,\prime}_{12} 
   \cdot \hat{\delta}_{r^{\,}_1r^{\,}_2} \, \hat{\vec{k}}^{\,}_{12}  
  \\
&
+ {\mathrm i} \, {W_0}  \, (\vec{\hat{\vsigma}}_1 + \vec{\hat{\vsigma}}_2 ) \cdot 
  \hat{\vec{k}}^{\,\prime}_{12} \, \times \, \hat{\delta}_{r^{\,}_1r^{\,}_2} 
  \hat{\vec{k}}^{\,}_{12} \, , \\
\hat{v}^{(3Sk)} 
&= 
{u_0} \, \Big( \hat{\delta}_{r_1r_3}\hat{\delta}_{r_2r_3} + \hat{\delta}_{r_3r_2}\hat{\delta}_{r_1r_2} + \hat{\delta}_{r_2r_1}\hat{\delta}_{r_3r_1} \Big)  \,, \\
\hat{v}^{(4Sk)} &= 
 {v_0} \, \Big( \hat{\delta}_{r_1r_3}\hat{\delta}_{r_2r_3} \hat{\delta}_{r_3r_4} 
 +\hat{\delta}_{r_1r_2}\hat{\delta}_{r_3r_2} \hat{\delta}_{r_2r_4} 
 +\cdots
 \Big) , \label{eq:Skyrme_int:pseudo_potential:4body}
\end{align}
\end{subequations}
where $t_i$, $x_i$, $W_0$, $u_0$ and $v_0$ are unknown
coupling constants, $\p \sigma12$ denotes the spin exchange 
operator, $\hat{\delta}_{r^{\,}_1r^{\,}_2}$ is the Dirac distribution, 
$\vec{\hat{\vsigma}}$ is the vector of Pauli spin matrices and 
$\hat{\vec{k}}^{\,2}_{12}$ and $\hat{\vec{k}}^{\,\prime \, 2}_{12}$ 
are the incoming and outgoing relative momenta. The three-body and
four-body terms contain three and twelve permutations 
of the coordinates, respectively.
The SR and MR energy kernels are calculated as
\begin{equation}
\label{eq:Intro_met:energyDFH_kernel}
\frac{\langle \Phi | \hat{H} | \Phi' \rangle}
     {\langle \Phi \vert \Phi' \rangle}
= \int \! d^3 r \; {\cal E}_{H}[\rho_{ij}^{\Phi \Phi'}, 
                                \kappa_{ij}^{\Phi \Phi'}, 
                                \kappa^{\Phi' \Phi \ast}_{ij}](\vec{r}) \, , \\
\end{equation}
where $\rho_{ij}^{\Phi \Phi'}$ and $\kappa_{ij}^{\Phi \Phi'}$ denote 
normal and anomalous transition density matrices, respectively. The resulting 
nuclear part of the energy density can be decomposed into bilinear, 
trilinear and quadrilinear parts according to their content in 
$\rho$ and $\kappa$
\begin{align}
{\cal E}_{H} 
&=
    {\cal E}^{\rho\rho}_{H}
+ {\cal E}^{\kappa\kappa}_{H}
+ {\cal E}^{\rho\rho\rho}_{H}
+ {\cal E}^{\kappa\kappa\rho}_{H}
\label{eq:Skyrme_int:energy_density} \nonumber
\\ & \hspace{1cm}
+ {\cal E}^{\rho\rho\rho\rho}_{H}
+ {\cal E}^{\kappa\kappa\rho\rho}_{H}
+ {\cal E}^{\kappa\kappa\kappa\kappa}_{H}
\, .
\end{align}
The bilinear parts take the usual form~\cite{Per04a}
\begin{subequations}
\begin{align}
\label{eq:Skyrme_int:2bodyEDF:normal}
{\cal E}_{H}^{\rho \rho} &= 
\sum_{q} \bigg\{ \nonumber
 A^{\rho_1 \rho_1} \rho_{q} \rho_{q}   
+ A^{\rho_1 \rho_2} \rho_{q} \rho_{\bar{q}}   
+ A^{s_1 s_1} \vec{s}_{q} \cdot \vec{s}_{q}   
 \\ &\nonumber 
+ A^{s_1 s_2} \vec{s}_{q} \cdot \vec{s}_{\bar{q}}   
+ A^{\tau_1 \rho_1} \tau_{q} \rho_{q}   
+ A^{\tau_1 \rho_2} \tau_{q} \rho_{\bar{q}}   
+ A^{T_1 s_1} \vec{T}_{q} \cdot \vec{s}_{q}   
 \\ &\nonumber 
+ A^{T_1 s_2} \vec{T}_{q} \cdot \vec{s}_{\bar{q}}   
+ A^{j_1 j_1} \vec{j}_{q} \cdot \vec{j}_{q}   
+ A^{j_1 j_2} \vec{j}_{q} \cdot \vec{j}_{\bar{q}}   
 \\ &\nonumber 
+ A^{\nabla \rho_1 \nabla \rho_1} \left[ \vnabla \rho_{q} \right] \cdot \left[ \vnabla \rho_{q}  \right] 
+ A^{\nabla \rho_1 \nabla \rho_2} \left[ \vnabla \rho_{q} \right] \cdot \left[ \vnabla \rho_{\bar{q}} \right]
 \\ &\nonumber 
+ \sum_{\mu \nu} \Big(
A^{\nabla s_1 \nabla s_1} \left[ \nabla_\mu s_{q, \nu} \right] \left[ \nabla_\mu s_{q, \nu}  \right] 
+ A^{J_1 J_1} J_{q, \mu \nu} J_{q, \mu \nu}   
 \\ &\nonumber 
+ A^{\nabla s_1 \nabla s_2} \left[ \nabla_\mu s_{q, \nu} \right] \left[ \nabla_\mu s_{\bar{q}, \nu} \right] 
+ A^{J_1 J_2} J_{q, \mu \nu} J_{\bar{q}, \mu \nu}   \Big)
 \\ &\nonumber 
+ A^{\rho_1 \nabla J_1} \rho_q \vnabla \cdot \vec{J}_q
+ A^{\rho_1 \nabla J_2} \rho_q \vnabla \cdot \vec{J}_{\bar{q}}
  \\ &        
+ A^{j_1 \nabla s_1} \vec{j}_q \cdot \vnabla \times \vec{s}_q 
+ A^{j_1 \nabla s_2} \vec{j}_q \cdot \vnabla \times \vec{s}_{\bar{q}}
\bigg\} \, , 
\\
 \label{eq:Skyrme_int:2bodyEDF:anormal}
{\cal E}_{H}^{\kappa \kappa} &= 
\sum_{q} \bigg\{ \nonumber
 A^{\tilde{\rho}^*_1 \tilde{\rho}_1} \tilde{\rho}^*_{q} \tilde{\rho}_{q}   
+ A^{\tilde{\tau}^*_1 \tilde{\rho}_1} \tilde{\tau}^*_{q} \tilde{\rho}_{q}   
+ A^{\tilde{\tau}_1 \tilde{\rho}^*_1} \tilde{\tau}_{q} \tilde{\rho}^*_{q}   
 \\ &\nonumber 
+ A^{\nabla \tilde{\rho}^*_1 \nabla \tilde{\rho}_1} \left[ \vnabla \tilde{\rho}^*_{q}  \right] \cdot \left[ \vnabla \tilde{\rho}_{q}  \right]
+ \sum_{\mu \nu} \Big(
A^{\tilde{J}^*_1 \tilde{J}_1}_1 \tilde{J}^*_{q, \mu \nu} \tilde{J}_{q, \mu \nu}   
 \\ &
+ A^{\tilde{J}^*_1 \tilde{J}_1}_2 \tilde{J}^*_{q, \mu \mu} \tilde{J}_{q, \nu \nu}   
+ A^{\tilde{J}^*_1 \tilde{J}_1}_3 \tilde{J}^*_{q, \mu \nu} \tilde{J}_{q, \nu \mu}    
\Big)
\bigg\}  \, .
\end{align}
\end{subequations}

\begin{table}[b!]
\begin{center}
\begin{tabular}{rccccccc}
\hline \hline \\[-3.7mm]
&$t_0$&$t_0x_0$&$t_1$&$t_1x_1$&$t_2$&$t_2x_2$&$W_0$\\
\hline\\[-2.5mm]
$A^{\rho_1 \rho_1}=$&$+\frac{1}{4}$&$-\frac{1}{4}$&&&&&\\[0.3mm]
$A^{\rho_1 \rho_2}=$&$+\frac{1}{2}$&$+\frac{1}{4}$&&&&&\\[0.3mm]
$A^{s_1 s_1}=$&$-\frac{1}{4}$&$+\frac{1}{4}$&&&&&\\[0.3mm]
$A^{s_1 s_2}=$&&$+\frac{1}{4}$&&&&&\\[0.3mm]
$A^{\tau_1 \rho_1}=$&&&$+\frac{1}{8}$&$-\frac{1}{8}$&$+\frac{3}{8}$&$+\frac{3}{8}$&\\[0.3mm]
$A^{\tau_1 \rho_2}=$&&&$+\frac{1}{4}$&$+\frac{1}{8}$&$+\frac{1}{4}$&$+\frac{1}{8}$&\\[0.3mm]
$A^{T_1 s_1}=$&&&$-\frac{1}{8}$&$+\frac{1}{8}$&$+\frac{1}{8}$&$+\frac{1}{8}$&\\[0.3mm]
$A^{T_1 s_2}=$&&&&$+\frac{1}{8}$&&$+\frac{1}{8}$&\\[0.3mm]
$A^{\nabla \rho_1 \nabla \rho_1}=$&&&$+\frac{3}{32}$&$-\frac{3}{32}$&$-\frac{3}{32}$&$-\frac{3}{32}$&\\[0.3mm]
$A^{\nabla \rho_1 \nabla \rho_2}=$&&&$+\frac{3}{16}$&$+\frac{3}{32}$&$-\frac{1}{16}$&$-\frac{1}{32}$&\\[0.3mm]
$A^{\nabla s_1 \nabla s_1}=$&&&$-\frac{3}{32}$&$+\frac{3}{32}$&$-\frac{1}{32}$&$-\frac{1}{32}$&\\[0.3mm]
$A^{\nabla s_1 \nabla s_2}=$&&&&$+\frac{3}{32}$&&$-\frac{1}{32}$&\\[0.3mm]
$A^{j_1 j_1}=$&&&$-\frac{1}{8}$&$+\frac{1}{8}$&$-\frac{3}{8}$&$-\frac{3}{8}$&\\[0.3mm]
$A^{j_1 j_2}=$&&&$-\frac{1}{4}$&$-\frac{1}{8}$&$-\frac{1}{4}$&$-\frac{1}{8}$&\\[0.3mm]
$A^{J_1 J_1}=$&&&$+\frac{1}{8}$&$-\frac{1}{8}$&$-\frac{1}{8}$&$-\frac{1}{8}$&\\[0.3mm]
$A^{J_1 J_2}=$&&&&$-\frac{1}{8}$&&$-\frac{1}{8}$&\\[0.3mm]
$A^{\rho_1 \nabla J_1}=$&&&&&&&$-1$\\[0.3mm]
$A^{\rho_1 \nabla J_2}=$&&&&&&&$-\frac{1}{2}$\\[0.3mm]
$A^{j_1 \nabla s_1}=$&&&&&&&$-1$\\[0.3mm]
$A^{j_1 \nabla s_2}=$&&&&&&&$-\frac{1}{2}$\\[1.3mm]
\hline\\[-3.0mm]
$A^{\tilde{\rho}^*_1 \tilde{\rho}_1}=$&$+\frac{1}{4}$&$-\frac{1}{4}$&&&&&\\[0.3mm]
$A^{\tilde{\tau}^*_1 \tilde{\rho}_1}=$&&&$+\frac{1}{8}$&$-\frac{1}{8}$&&&\\[0.3mm]
$A^{\tilde{\tau}_1 \tilde{\rho}^*_1}=$&&&$+\frac{1}{8}$&$-\frac{1}{8}$&&&\\[0.3mm]
$A^{\nabla \tilde{\rho}^*_1 \nabla \tilde{\rho}_1}=$&&&$+\frac{1}{16}$&$-\frac{1}{16}$&&&\\[0.3mm]
$A^{\tilde{J}^*_1 \tilde{J}_1}_1=$&&&&&$+\frac{1}{4}$&$+\frac{1}{4}$&\\[0.3mm]
$A^{\tilde{J}^*_1 \tilde{J}_1}_2=$&&&&&&&$+\frac{1}{2}$\\[0.3mm]
$A^{\tilde{J}^*_1 \tilde{J}_1}_3=$&&&&&&&$-\frac{1}{2}$\\[1.5mm]
\hline \hline
\end{tabular}
\begin{tabular}{rcc c|c|c rcc}
&$u_0$&$v_0$& &&& &$u_0$&$v_0$ \\
\hline
&&&&&&&&
\\[-2.5mm]
$A^{\rho_1 \rho_1 \rho_2}=$&$+\frac{3}{4}$&&\quad &\quad &\quad &$A^{\tilde{\rho}^*_1 \tilde{\rho}_1 \rho_2}=$&$+\frac{3}{4}$&\\[0.3mm]
$A^{s_1 s_1 \rho_2}=$&$-\frac{3}{4}$&&\quad &\quad &\quad &$A^{\tilde{\rho}_1\tilde{\rho}_1\tilde{\rho}_2\tilde{\rho}_2}=$&&$+\frac{3}{8}$\\[0.3mm]
$A^{\rho_1\rho_1\rho_2\rho_2}=$&&$+\frac{3}{8}$&\quad &\quad &\quad &$A^{\tilde{\rho}_1\tilde{\rho}_1\rho_2\rho_2}=$&&$+\frac{3}{4}$\\[0.3mm]
$A^{s_1 s_1 s_2 s_2}=$&&$+\frac{3}{8}$&\quad &\quad &\quad &$A^{\tilde{\rho}_1\tilde{\rho}_1 s_2 s_2}=$&&$-\frac{3}{4}$\\[0.3mm]
$A^{\rho_1 \rho_1 s_2 s_2}=$&&$-\frac{3}{4}$&&&&&&\\[1.5mm]
\hline \hline
\end{tabular}
\end{center}
\caption{
\label{tab:Skyrme_int:EDF:coeff}
Coupling constants of the EDF expressed in terms of the 
parameters of the pseudo-potential.
Missing entries are zero.
}  
\end{table}

\noindent
They are functions of local matter density $\rho_q(\vec{r})$, kinetic matter 
density $\tau_q(\vec{r})$, current density $\vec{j}_{q}(\vec{r})$, spin 
density $\vec{s}_{q}(\vec{r})$, spin-kinetic density $\vec{T}_{q}(\vec{r})$, 
spin-current density $J_{q,\mu\nu}(\vec{r})$, pairing density 
$\tilde{\rho}_{q}(\vec{r})$, pairing-kinetic density $\tilde{\tau}_q(\vec{r})$
and pairing-current density $\tilde{J}_{q,\mu\nu}(\vec{r})$~\cite{Per04a}.
Here, we choose a proton-neutron representation where the sum runs over 
neutron and proton densities \mbox{$q = n$}, $p$, and where 
\mbox{$\bar{q} \neq q$} denotes the other nucleon species. The trilinear 
parts of the energy density read
\begin{subequations}
\begin{align}
{\cal E}^{\rho \rho \rho}_{H} &= \sum_{q}
\Big\{
   A^{\rho_1 \rho_1 \rho_2} \rho_{q} \rho_{q} \rho_{\bar{q}}   
+A^{s_1 s_1 \rho_2} \vec{s}_{q} \cdot \vec{s}_{q} \rho_{\bar{q}}   
\Big\}  \label{eq:Skyrme_int:3bodyEDF:normal}
\,,
\\
{\cal E}^{\kappa \kappa \rho}_{H} &= \sum_{q}
A^{\tilde{\rho}^*_1 \tilde{\rho}_1 \rho_2} \tilde{\rho}^*_{q} \tilde{\rho}_{q} \rho_{\bar{q}}   
\, , \label{eq:Skyrme_int:3bodyEDF:anormal}
\end{align}
\end{subequations}
whereas the quadrilinear parts are given by
\begin{subequations}
\begin{align}
 \label{eq:Skyrme_int:4bodyEDF:normal}
{\cal E}^{\rho \rho \rho \rho}_{H} &= \sum_{q} 
\Big\{
A^{\rho_1\rho_1\rho_2\rho_2} \rho_q \rho_q \rho_{\bar{q}} \rho_{\bar{q}}
+A^{s_1 s_1 s_2 s_2}  \vec{s}_q \cdot \vec{s}_q  \vec{s}_{\bar{q}} \cdot \vec{s}_{\bar{q}}
\nonumber \\ & \hspace{1cm}
+A^{\rho_1 \rho_1 s_2 s_2} \rho_q \rho_q \vec{s}_{\bar{q}} \cdot \vec{s}_{\bar{q}}
\Big\}
\, ,
\\
\label{eq:Skyrme_int:4bodyEDF:anormal1}
{\cal E}^{\kappa \kappa \rho \rho}_{H} &= 
\sum_{q} 
\Big\{
A^{\tilde{\rho}_1\tilde{\rho}_1\rho_2\rho_2} \tilde{\rho}_q^*\tilde{\rho}_q^{\,} \rho_{\bar{q}} \rho_{\bar{q}}
+A^{\tilde{\rho}_1\tilde{\rho}_1 s_2 s_2} \tilde{\rho}_q^*\tilde{\rho}_q^{\,} \vec{s}_{\bar{q}} \cdot \vec{s}_{\bar{q}}
\Big\}
\,,
\\
 \label{eq:Skyrme_int:4bodyEDF:anormal2}
{\cal E}^{\kappa \kappa \kappa \kappa}_{H} &= \sum_{q} 
A^{\tilde{\rho}_1\tilde{\rho}_1\tilde{\rho}_2\tilde{\rho}_2} 
\tilde{\rho}_q^*\tilde{\rho}_q^{\,} \tilde{\rho}_{\bar{q}}^*\tilde{\rho}_{\bar{q}}^{\,}
\, .
\end{align}
\end{subequations}
Each term in the functional is provided with a coefficient whose 
superscripts are labelled by numbers $1$ and $2$ that represent $q$ 
and $\bar{q}$, respectively. The coefficients are related to 
the parameters of the pseudo-potential~(\ref{eq:Skyrme_int:pseudo_potential}) 
through Table~\ref{tab:Skyrme_int:EDF:coeff}.

\section{A tentative fit}

>From the published work on pseudo-potential-based Skyrme EDFs from 
the 1970s it is clear from the outset that the simple form of
Eq.~(\ref{eq:Skyrme_int:hamiltonian}) will be unlikely 
to reach the quality of modern parametrisations of the general 
Skyrme EDF. Our more modest aim is a parametrisation that could be
used in time-reversal-breaking SR and angular-momentum 
restored MR EDF calculations based on HFB-type reference states. 
To accomplish this, we had to add several new constraints to our
fit protocol. First, we had to avoid any unphysical instabilities, 
not just the ones related to too small values of the Landau parameters, 
but also finite-size instabilities of the kind discussed in 
Refs.~\cite{lesinski06,pastore}. Second, we aimed at an overall 
repulsive interaction in the spin channels and at an attractive pairing
interaction that provides gaps of realistic size. This, however,
required to relax the constraints on many other properties, in 
particular those of nuclear matter.

\begin{table}[!t]
\begin{center}
\begin{tabular}{cc r}
\hline \hline \\[-3.7mm]
Param. & & SLyMR0 \qquad \\
\hline 
$t_0$& (MeV.fm$^3$) \quad & \quad  \;$-1210.093228$\; \\
$x_0$&  \quad & \quad  \;$-0.283818$\; \\
$t_1$& (MeV.fm$^5$)  \quad & \quad  \;$632.460114$\; \\
$x_1$&  \quad & \quad \;$-0.038032$\; \\
$t_2$ & (MeV.fm$^5$)  \quad & \quad  \;$45.081296$\; \\
$x_2$&  \quad & \quad  \;$1.849978$\; \\
$W_0$& (MeV.fm$^5$)  \quad & \quad  \;$122.618466$\; \\
$u_0$& (MeV.fm$^6$)  \quad & \quad  \;$2529.151191$\; \\
$v_0$& (MeV.fm$^9$)  \quad & \quad  \;$-14750.0$\; \\[1.5mm]
\hline \hline 
\end{tabular}
\end{center}
\caption{
\label{tab:Skyrme_int:param}
Coupling constants of SLyMR0.
}
\end{table}

\begin{table}[b!]
\begin{center}
\begin{tabular}{rcccccc}
\hline \hline \\[-3.7mm]
Param. & $\rho_{\text{sat}}$ & $E/A$ & $a_{\text{sym}}$ & $m^*_0/m$ & $K_{\infty}$ & $g_0$\\
&(fm$^{-3}$) &(MeV) & (MeV) &  & (MeV) &   \\
\hline\\[-2.5mm]
SIII  \cite{beiner75}  &$0.145$&$-15.85$&$28.2$&$0.76$&$355.3$&$-1.58$\\[1.5mm]
SIV   \cite{beiner75}  &$0.151$&$-15.96$&$31.2$&$0.47$&$324.6$&$0.06$\\[1.5mm]
SV    \cite{beiner75}  &$0.155$&$-16.05$&$32.8$&$0.38$&$305.7$&$0.57$\\[1.5mm]
SHZ2  \cite{Sat12a}    &$0.157$&$-16.27$&$42.1$&$0.38$&$309.6$&$0.27$\\[1.5mm]
SLyMR0                 &$0.152$&$-15.04$&$23.0$&$0.47$&$264.2$&$0.88$\\[1.5mm]
SLy4  \cite{chabanat97}&$0.160$&$-15.97$&$32.0$&$0.69$&$229.9$&$1.38$\\[1.5mm]
\hline \hline
\end{tabular}
\end{center}
\caption{
\label{tab:Skyrme_int:paramINM}
Saturation density $\rho_{\text{sat}}$, energy per particle $E/A$, 
symmetry energy $a_{\text{sym}}$, effective mass $m^*_0$, 
incompressibility $K_{\infty}$ and spin-Landau parameter $g_0$ at 
saturation for the parametrisations studied here.
}  
\end{table}

To construct an acceptable starting point for a fit, we took the 
parameters of SIV, which has a reasonable value of $g_0$ 
and provides weakly attractive pairing, from 
Ref.~\cite{beiner75}. We modified $x_1$ and $t_1$ 
to enhance pairing and then $u_0$, $v_0$ and, to a lesser extend, 
$t_2$ and $x_2$, to bring the parametrisation back to a better
description of symmetric INM. Simultaneously,
$x_0$ had to be adapted to reject the onset of spin instabilities 
far enough above the saturation density. In a second step, the parameters
were then fine-tuned to describe a set of nuclear masses and radii 
in addition to the nuclear matter properties in a least-square fit.
There is no "best fit" in the usual sense, however. As several 
nuclear matter properties remain far from the empirical values, 
there are many possible fits of very similar 
(limited) quality. A typical parameter set obtained with that procedure 
that we will use for further studies is given in 
Tab.~\ref{tab:Skyrme_int:param}. It is called SLyMR0 in 
what follows.

Table~\ref{tab:Skyrme_int:paramINM} compares some of the associated 
nuclear matter properties with those of other parametrisations
used as pseudo-potentials
in the literature. They differ in higher-order terms in 
Eq.~(\ref{eq:Skyrme_int:pseudo_potential}). 
SV and SHZ2 are pure two-body forces that have been used 
for isospin and angular-momentum projection of HF states (without 
pairing)~\cite{Sat10a,Sat12a}. SIII and SIV include 
the three-body term, whereas SLyMR0 contains three- and 
four-body terms. Values obtained with a standard parametrisation 
SLy4 of the general Skyrme EDF are given for comparison.

As expected, all the pseudo-potential-based parametrisations
have difficulties to describe nuclear matter properties.
In particular, their effective mass is very low except for
SIII. This leads to a density of single-particle levels
around the Fermi energy that is much lower than the empirical one.
For SIII, the larger effective mass leads, through the
interrelations between the coupling constants, to spin instability
as indicated by $g_0 < -1$ \cite{Bac75a}. For SIV, the reduction of 
the effective mass within the same functional form pushes $g_0$ 
to values around zero. However, an analysis along the lines of 
Ref.~\cite{pastore} reveals that SIV has a finite-size spin
instability in the \mbox{$S=1$}, \mbox{$T=0$} channel that cannot
be revealed by the Landau parameters. Hence, neither SIII nor
SIV can be used in their pseudo-potential form in time-reversal
invariance breaking mean-field calculations. By contrast, 
such an analysis does not indicate unphysical instabilities for 
SV, SHZ2 and SLyMR0 at densities relevant for low-energy nuclear 
structure.

\begin{figure}[t!]
\begin{center}
\includegraphics[width=\linewidth]{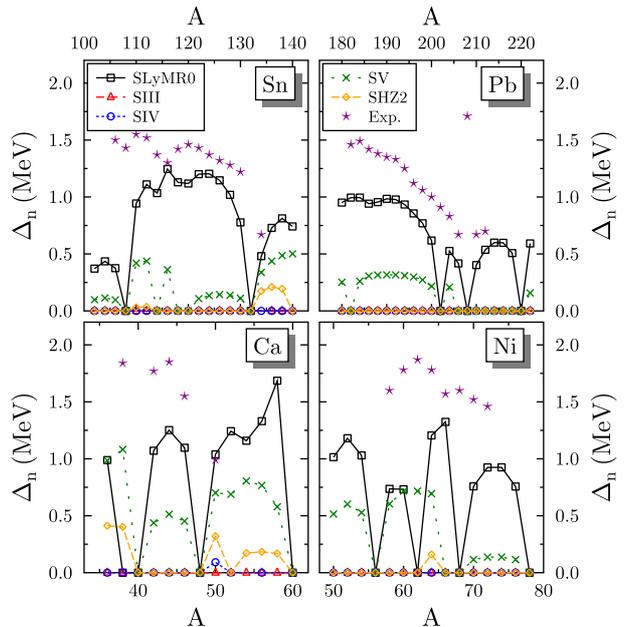}\\
\vspace{-6mm}
\caption{
\label{fig:Skyrme_res:gaps}
Neutron spectral gaps of singly magic even-even nuclei in
the isotopic chains as indicated at the SR-EDF level. }
\end{center}
\end{figure}

In Fig.~\ref{fig:Skyrme_res:gaps}, pairing properties
are examined via the spectral gap 
$E_{\text{pair},n} / \int \! d^3r \; \tilde{\rho}_n(\vec{r})$
of neutrons as obtained from spherical HFB-type SR calculations of 
singly magic even-even nuclei. Empirical pairing gaps obtained 
from a three-point mass difference are shown for comparison. When 
solving HFB equations, pairing matrix elements have been 
multiplied with a smooth cutoff at $8.5$~MeV above and below the 
Fermi energy. 
SIII, SIV, SHZ2 and SV give null pairing or at least a weak pairing in some nuclei. 
Only SLyMR0, for which this property was enforced 
during the fit, provides pairing gaps of a realistic size.

\begin{figure}[t!]
\begin{center}
\includegraphics[width=\linewidth]{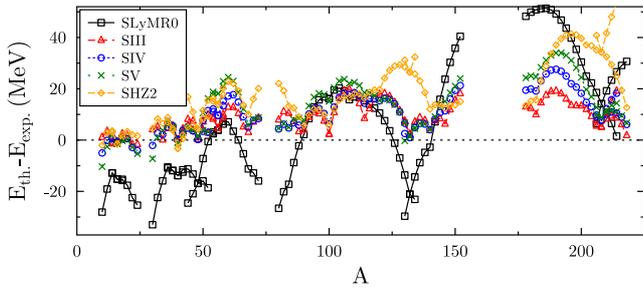}\\
\vspace{-6mm}
\caption{
\label{fig:Skyrme_res:E}
Binding energy residuals as a function of $A$ for
singly magic nuclei for spherical SR-EDF calculations. 
Nuclei in isotonic and isotopic chains are connected 
by lines.
}
\end{center}
\end{figure}

Figure~\ref{fig:Skyrme_res:E} exhibits mass residuals for 
isotopic and isotonic chains of singly magic nuclei as 
obtained from spherical HFB-type SR calculations. The particularly 
large drift of the curves obtained with SLyMR0 results mainly from 
its very low value for the asymmetry energy coefficient 
$a_{\text{sym}}$, which cannot be increased without jeopardising 
pairing or the time-odd terms in the EDF. We have checked that, 
in spite of its poor description of masses, SLyMR0 gives a reasonable 
description of the deformation of nuclei in the $sd$ and $pf$ 
shell regions, and of their rotational bands in cranked HFB
calculations.

\section{Conclusion}

The simple pseudo-potential~(\ref{eq:Skyrme_int:pseudo_potential}) 
allows for a description of basic nuclear properties that does 
not meet the standard of state-of-the-art SR-EDF calculations. 
However, it is sufficient for its main purpose of benchmarking 
multi-dimensional MR EDF calculations, which then will be free 
from the pathologies encountered with any modern standard parametrisation. 
In particular, we achieved 
a fair description of pair correlations while avoiding (finite-size)
spin-instabilities. However, the simple form used here is clearly
over-constrained and higher-order terms will be needed to replicate
the performance of the best available general Skyrme EDFs at the
SR level. Work in that direction is underway. As a first step, the
most general central three-body contact operator 
up to second order in gradients has been worked out 
recently~\cite{sadoudithesis}. 
First fits that aim at the set-up of a suitable protocol are currently
underway and show promising results.

\begin{ack}

Clarifying discussions with B.~Avez, J.~Meyer and A.~Pastore 
are gratefully acknowledged.
This work has been supported by the the Agence Nationale 
de la Recherche under Grant No.~ANR 2010 BLANC 0407 "NESQ".

\end{ack}


\end{document}